\documentclass[final,epsfig]{siamltex}
\usepackage{epsfig}

\newcommand{\eeq}{\end{equation}}
\newcommand{\beq}{\begin{equation}}
\newcommand{\nuq}[1]{\label{#1} \eeq}
\newcommand{\ba}{\begin{array}}
\newcommand{\ea}{\end{array}}

\newcommand{\bino}[2]{\mbox{$(\stackrel{#1}{\scriptstyle #2})$}}

\newcommand{\ee}{\end{equation}}
\newcommand{\be}{\begin{equation}}
\newtheorem{prop}{Proposition}
\newcommand{\epr}{\end{prop}}
\newcommand{\bpr}{\begin{prop}}
\newtheorem{teo}{Theorem}
\newcommand{\bth}{\begin{teo}}
\newcommand{\eth}{\end{teo}}
\newtheorem{rema}{Remark}
\newcommand{\bre}{\begin{rema}}
\newcommand{\ere}{\end{rema}}

\title{Fourier--Bessel functions of singular continuous measures
and their many asymptotics
\thanks{Dedicated to Ed Saff for his Sixtieth.}}

\author{Giorgio Mantica\thanks{Center for Non-linear and Complex Systems,
Universit\`a dell'Insubria, Via Valleggio 11, 22100 Como, Italy
({\tt giorgio@uninsubria.it}).}
        }

\begin{document}

\maketitle

\begin{abstract}
We study the Fourier transform of polynomials in an orthogonal
family, taken with respect to the orthogonality measure. Mastering
the asymptotic properties of these transforms, that we call
Fourier--Bessel functions, in the argument, the order, and in
certain combinations of the two is required to solve a number of
problems arising in quantum mechanics. We present known results,
new approaches and open conjectures, hoping to justify our belief
that the importance of these investigations extends beyond the
application just mentioned, and may involve interesting
discoveries.
\end{abstract}

\begin{keywords}
Singular measures, Fourier transform, orthogonal polynomials,
almost periodic Jacobi matrices, Fourier-Bessel functions, quantum
intermittency, Julia sets, iterated function systems, generalized
dimensions, potential theory.
\end{keywords}

\begin{AMS}
42C05, 33E20, 28A80, 30E15, 30E20
\end{AMS}

\pagestyle{myheadings} \thispagestyle{plain} \markboth{G.
MANTICA}{FOURIER--BESSEL FUNCTIONS OF SINGULAR CONTINUOUS
MEASURES}

\section{Introduction and examples}

\begin{figure}[ht]
%  \vspace{2.5in}
% \centerline{\epsfxsize=3in,\epsffile{gfunz2.ps}}
\includegraphics[width=8cm,height=12cm,angle=270]{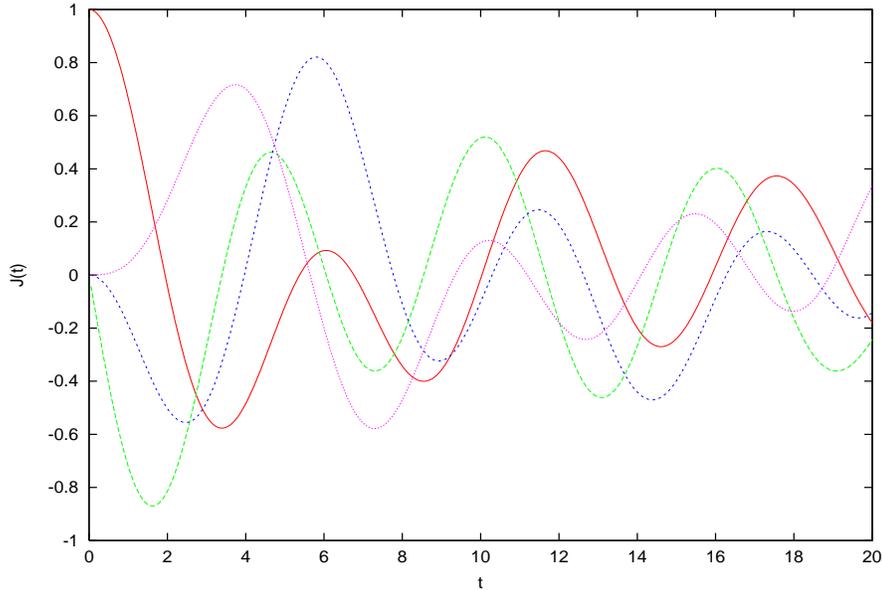}
\caption{F-B. functions $i^n {\cal J}_n(\mu;t)$, $n=0,1,2,3$, for
a Julia set measure with $\lambda = 2.9$. Different curves can be
distinguished from the behavior at the origin: ${\cal J}_n(\mu;t)
\sim t^n$, as in the Bessel case. } \label{fbes1}
\end{figure}

Let $\mu$ be a positive measure, for which the moment problem is
determined, and let $\{p_n(\mu;s)\}_{n \in {\bf N}}$ be its
orthogonal polynomials. The {\em Fourier-Bessel functions} (F-B.
for short) ${\cal J}_n(\mu;t)$ are the Fourier transforms of
$p_n(\mu;s)$ with respect to $\mu$:
 \beq
   {\cal J} _n(\mu;t) :=
     \int d\mu(s) ~p_n(\mu;s)~e^{-its}.
 \nuq{geb1}
This nomenclature follows---for lack of better candidates---from
the simple observation that when $\mu$ is the continuous measure
with density $d\mu(s) = \frac{ds}{\pi \sqrt{1-s^2}}$, and
therefore $p_n(\mu;s)$ are the (properly normalized) Chebyshev
polynomials, the F-B. functions are the usual integer order Bessel
functions: ${\cal J} _n(\mu;t) = (-i)^n J_n(t)$. When the measure
is symmetrical with respect to the origin,  as in this case, the
F-B. functions are either real, or purely imaginary. A graph of
the first few F-B. functions, multiplied  by $i^n$,  is displayed
in Fig. \ref{fbes1} for a singular continuous measure supported on
a real Julia set (to be introduced in the following). Notice the
joyful oscillations that these F-B. functions feature, as opposed
to the more disciplined, and in the end boring attitude of the
$J_n$'s. This paper wants to be an ode to the fascinating
properties of F-B. functions of singular continuous measures, that
in my opinion are still largely unexplored: I shall present a few
results, but mostly open problems. The style of this paper will be
suggestive of possible developments, rather than assertive of
formal results, and at times I shall gladly renounce to rigor in
favor of intuition, hoping with confidence that others will take
up where I have left, and complete the picture. In this way, I
believe to be correctly interpreting Ed's attitude towards
mathematics as a communal endeavor, and it is not only a pleasure
for me, but an honor, to dedicate to him these notes.

\begin{figure}[ht]
\includegraphics[height=12cm,width=8cm,angle=270]{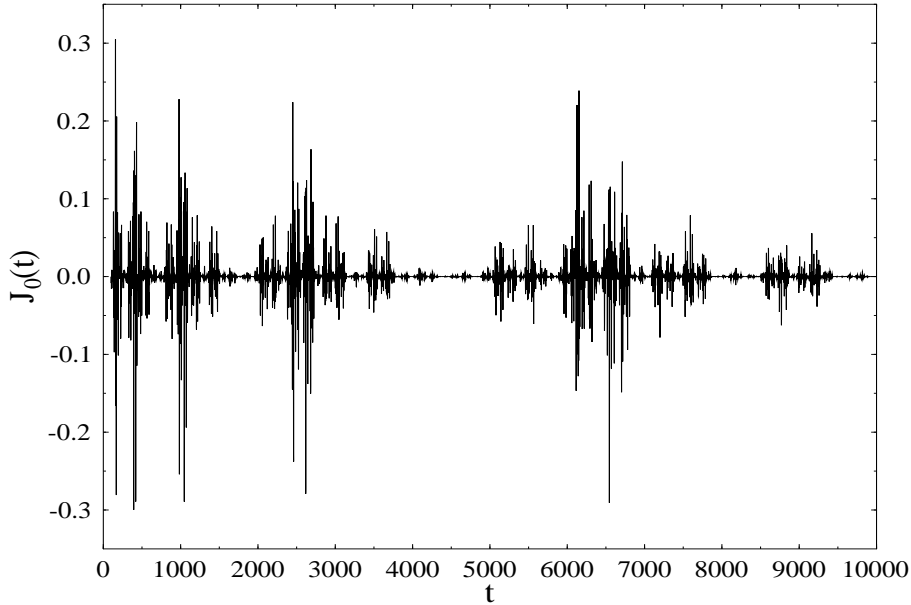}
\caption{F-B. function ${\cal J}_0(\mu;t)$ for an I.F.S. measure,
% described in Sect. \ref{ifssec},
over a larger $t$-scale than in Fig. (\ref{fbes1}).} \label{fbes2}
\end{figure}

The asymptotic of F-B. functions for large values of the argument,
$t$, is a classical theme of investigation, especially when $n=0$,
since ${\cal J}_0(\mu;t)$ is the Fourier transform of the measure
$\mu$ \cite{str0,str1,str2,maka,lw1}. In this study, the nature of
the orthogonality measure $\mu$ plays a major r\^ole. In fact, it
is in the realm of singular, {\em multi-fractal} measures that the
most interesting phenomena appear. First of all, at difference
with the usual Bessel case, convergence of ${\cal J} _n(\mu;t)$ to
zero is not to be expected, and indeed in Figure \ref{fbes2}, that
depicts a much larger argument range than Fig. \ref{fbes1}, this
time for a measure associated with a linear Iterated Function
System, bursts of ``activity'' of ${\cal J}_0(\mu;t)$ are
observed, amidst zones of quiescence. Because of similarities with
the theory of turbulence,
% that I have explained elsewhere ,
I have termed this phenomenon and its consequences {\em quantum
intermittency} \cite{papgua,physd1,physd2}.

A common technique to cope with these bursts is to take suitable
time averages, like Cesaro's. After averaging, decay of ${\cal
J}_n(\mu;t)$ to zero actually takes place, according to an
algebraic law. Now, two main problems can be investigated: the
decay of the averaged F-B. functions themselves, and that of their
(averaged) square moduli, this second problem having received
larger attention than the first. In two recent papers
\cite{pagen,qagen} we have collected known and new results on
these questions, under the unifying theme of Mellin transforms.
The following scheme is encountered in these theorems, under very
broad hypotheses (typically, the existence of orthogonal
polynomials): for any $x$ less than the divergence abscissa of a
potential theoretic function, the Cesaro average of F-B. functions
(or of their square moduli) decays faster than $t^{-x}$. The
divergence abscissas entering these theorems are identified as the
local dimension of the measure at zero in the first case, and as
the correlation dimension of the measure in the second.
The appearance of dimensional quantities of the orthogonality
measure is not accidental: indeed, they play a major r\^ole in the
asymptotics of F-B. functions, as it will become apparent in the
following.

Quite different is the asymptotic behavior for large values of the
order, and fixed argument.  A general result can be obtained on
the basis of a Chebyshev expansion of the matrix exponential
\cite{oberw}: this theorem states that under the sole hypothesis
that the support of $\mu$ is bounded, at fixed time $t$, for any
$\alpha \geq 0$, there exist a constant $C_\alpha$ so that the
F-B. functions ${\cal J}_n(\mu;t)$ decay faster than exponentially
in $n$:
\begin{equation}
   |{\cal J}_n(\mu;t) | \leq C_\alpha e^{-\alpha n}
  \;\; \mbox{for all} \; n.
\label{tev2}
\end{equation}
The need to refine this estimate will become apparent in Sect.
\ref{tailsec}.

So far we have described asymptotic questions of a quite
conventional kinship. The best way to introduce and motivate the
new questions that we would like to answer,  is to outline a
quantum mechanical interpretation of the F-B. functions. An
alternative physical interpretation, that considers the
propagation of excitations in chains of classical linear
oscillators, can be found in \cite{physd2}.

Recall that the orthogonal polynomials $\{p_n(\mu;s)\}_{n \in {\bf
N}}$ satisfy a recursion relation that can be written in vector
form as
 \beq
   s \; {\bf p}(\mu;s) =  {\bf J}_\mu  {\bf p}(\mu;s),
 \nuq{opol2}
 % \beq
   % sp_n(\mu;s) = a_{n+1} p_{n+1}(\mu;s) +b_n  p_n(\mu;s) +a_n  p_{n-1}(\mu;s),
 % \nuq{opol2}
where
% $a_n>0$, and $b_n$ are the entries of
${\bf p}(\mu;s)$ is the infinite vector of orthogonal polynomials
evaluated at position $s$, and  ${\bf J}_\mu$ is the Jacobi matrix
uniquely associated with $\mu$ (in the case when the moment
problem is determined, of course). We can formally think of ${\bf
J}_\mu$ as a self-adjoint operator acting in the space of square
summable sequences, $l_2({\bf Z}_+)$ (for the precise treatment of
this part see \cite{qagen}), and consider the  evolution  that it
generates via Schr\"odinger equation:
 \beq
 i \frac{d}{dt} \psi(t) = {\bf J}_\mu \psi(t) .
 \nuq{sch}
In this equation, $\psi(t)$   is the {\em wave-function}, a vector
that evolves in the space $l_2({\bf Z}_+)$ and defines the state
of the quantum system. At any time $t$, we can compute the
projection of $\psi(t)$ on $e_n$, the $n$-th vector of the
canonical basis of $l_2({\bf Z}_+)$:
 \beq
   \psi_n(t) := (\psi(t),e_n),
   \nuq{psin}
    where $(\cdot,\cdot)$ denotes the scalar product in $l_2({\bf Z}_+)$.

The initial state of the evolution, $\psi(0)$, can be chosen
freely. Letting it coincide with the first basis vector, $e_0$,
leads to the conclusion \cite{qagen} that  $\psi_n(t)$, the
projection of the time evolution on the $n$-th basis state, {\em
can be precisely identified with} ${\cal J}_n(\mu;t)$, the $n$-th
F-B. function:
 \beq
   \psi_n(t) = {\cal J}_n(\mu;t).
   \nuq{spcthe}

The physical amplitudes of the quantum motion are the square
moduli of the projections of the wave-function on the basis states
of Hilbert space, $|\psi_n(t)|^2$. They are interpreted as the
quantum probability to find the system in the state $e_n$ at the
time $t$. As such, they can be used to define the expected values
of dynamical operators. Unitarity of the quantum evolution
operator, $e^{-it{\bf J}_\mu}$, implies the probability
conservation formula
 \beq
\sum_{n=0}^\infty |{\cal J}_n(\mu;t)|^2 = 1,
 \nuq{norma}
valid for all times $t$. This formula gives a new meaning to the
analogous one already known for integer order Bessel functions.

\begin{figure}[ht]
\includegraphics[width=8cm,height=14cm,angle=270]{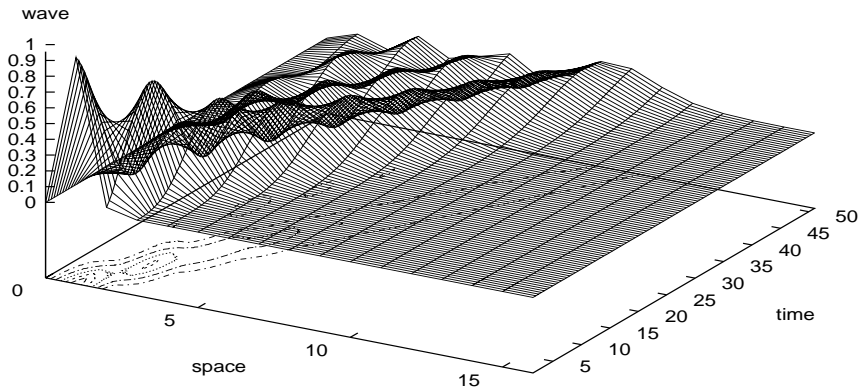}
\caption{F-B. functions $|{\cal J} _n(\mu;t)|^2$ versus time $t$
and space $n+1$, for $d\mu(s) = \frac{ds}{\pi \sqrt{1-s^2}}$.This,
and all three--dimensional graphs are zoomable for better
viewing.}

\label{wave2}
\end{figure}

\begin{figure}[ht]
\includegraphics[width=8cm,height=14cm,angle=270]{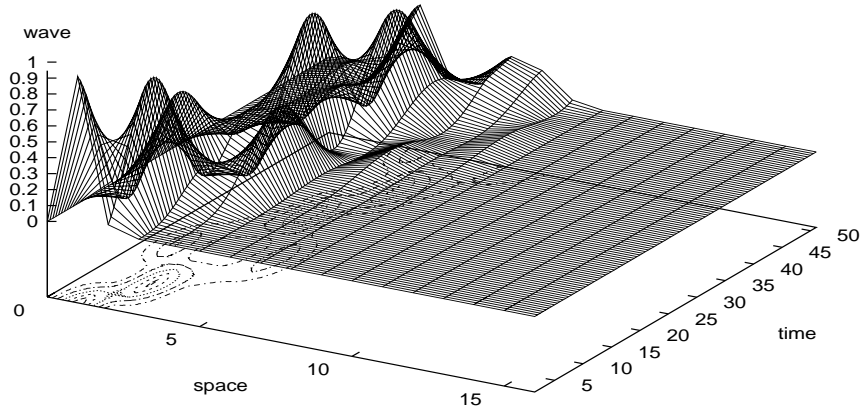}
\caption{F-B. functions $|{\cal J} _n(\mu;t)|^2$ versus time $t$
and space $n+1$, for a Julia set measure with $\lambda = 2.9$.}
\label{wave4}
\end{figure}

 Think now of $n$ as labelling the position in a
regular one-dimensional lattice. Then, $\psi(0)=e_0$ describes a
quantum system initially localized in the origin of this lattice,
and consequently $|\psi_n(t)|^2$ describes the {\em spreading} of
the quantum wave over this space. Figure \ref{wave2} shows the
initial part of the evolution in the case of the usual Bessel
functions (for which the measure $\mu$ is absolutely continuous),
and Figure \ref{wave4} displays the same information in the case
of a singular continuous Julia set measure. Differences between
the two are apparent.

 To gauge this spreading we utilize
the moments of the position $n$,
 \beq
  \nu_\alpha(t)
    := \sum_{n=0}^\infty n^\alpha |\psi_n(t)|^2 =
      \sum_{n=0}^\infty n^\alpha |{\cal J} _n(\mu;t)|^2.
      \nuq{anu1}
Here, the index $\alpha$ takes all positive real values
\cite{negamome}.
%For $\alpha=0$, the unitarity of quantum
%evolution implies that $\nu_0(t)=1$ for all values of $t$.

As it happens, for the singular measures that we are interested
in, the asymptotic behavior of the {\em position moments}
$\nu_{\alpha}(t)$ is power-law, with non--trivial exponents: we
therefore define the {\em growth exponents} $\beta^\pm(\alpha)$
via the upper and lower limits
 \beq
  \beta^{\pm} (\alpha) = \frac{1}{\alpha}
  \lim_{t \rightarrow \infty}
  \bino{sup}{inf}
  \frac{\log \nu_\alpha (t)}{\log t}.
 \nuq{beta1}
The functions $\beta^{\pm} (\alpha)$ are also called the {\em
quantum intermittency functions}.

In the setting so defined, trivially $\beta^{\pm} (\alpha) \leq
1$, and $\beta^{\pm} (\alpha) = 1$ in the Bessel case. For
singular measures, bounds related to dimensional characteristics
become of importance \cite{gsb}: under the sole request of
existence of the orthogonal polynomials of $\mu$, it is proven
that
 $
  \beta^-(\alpha) \geq \mbox{dim}_H (\mu),
$ where $\mbox{dim}_H (\mu)$ is the Hausdorff dimension of $\mu$,
and
$
   \beta^+(\alpha) \geq \mbox{dim}_p (\mu),
  $
the last quantity being the packing (or Tricot) fractal dimension.
Indeed, these theorems are even more general than required for our
purpose: they apply to any quantum evolution in a separable
Hilbert space, see the original references for details.

Notice that the above bounds do not depend upon the index
$\alpha$. According to my definition, {\em quantum intermittency}
is present when $\beta^{\pm} (\alpha)$ are {\em not} constant
functions of the argument $\alpha$. However strange it might seem
at first, this case is typical of singular continuous measures
supported on Cantor sets. The {\em name of the game} of much
recent theoretical research has therefore been to study these
functions, and to track the origin of their behavior in the
properties of the measure $\mu$, and of its orthogonal
polynomials. This is the problem that will be discussed in this
paper.

\section{Kinematics, and expansion in orthogonal polynomials}

The quantities described in the Introduction can be obviously
expressed in terms of orthogonal polynomials. In fact, the
position moments $\nu_{\alpha}(t)$ can be written as
 \beq
 \nu_{\alpha}(t) =\sum_{n=0}^{\infty} n^\alpha
 \int\!\!\int d\mu(s)d\mu(r) e^{i(r-s)t}
 p_{n}(\mu;s)p_{n}(\mu;r)\;.
\label{gal2}
 \eeq
We are therefore confronted with the highly singular kernel
 \beq
  K^\alpha_\mu(r,s) := \sum_{n=0}^{\infty} n^\alpha
  p_{n}(\mu;s)p_{n}(\mu;r).
 \nuq{kern}
 When $\alpha = 0$, we obtain the
 reproducing kernel of the orthogonal polynomials of $\mu$:
 $K^0_\mu(r,s) = \delta_\mu(r-s)$.

The behavior of individual F-B. functions can be rather erratic.
The common procedure is then to perform a time average. Cesaro
averaging is a common choice, but other forms of averaging work as
well. For instance, Gaussian averaging,
 \beq
 {\cal A}_G (f) (t) :=
  \frac{1}{2 t \sqrt{\pi} } \int_{-\infty}^\infty e^{-\frac{s^2}{t^2}} f(s) ds ,
     \nuq{ces2}
where $f$ is either $\nu_{\alpha}(t)$, or ${\cal J} _n(\mu;t)$,
has the advantage of a better regularity in the windowing
function: we have in fact
 \beq
 {\cal A}_G(\nu_{\alpha})(t) =\sum_{n=0}^{\infty} n^\alpha
 \int\!\!\int d\mu(s)d\mu(r) \chi_{1/t}(r-s)
 p_{n}(\mu;s)p_{n}(\mu;r),\;
\label{gal22}
 \eeq
where $\chi_{\omega}(u)=e^{-\frac{u^2}{\omega^2}}$ is a smooth
analogue of the characteristic function of the interval
$[-\omega,\omega]$. For ease of notation, we  use the convention
$\omega := t^{-1}$ throughout this paper.

\section{Distribution functions and lower bounds to the growth exponents}
\label{sshape} In the study of the general problem (\ref{beta1})
the consideration of a finite truncation of the   $\alpha=0$
moment, turns out to be useful. Define
 \beq
   \nu_0(N,\omega):=
   % \sum_{n=0}^{N} {\cal A}_G(|{\cal J}_n(\mu;t)|^2) =
  \int\!\!\int
     d\mu(s)\,d\mu(r)
     \chi_{\omega}(r-s)
%   e^{-\frac{(r-s)^2}{\omega^2}}
   \sum_{n=0}^{N}\, p_{n}(\mu;s)\,p_{n}(\mu;r)\;.
   \label{gal4}
\eeq
 This is the Gaussian time average, up to time $t=\omega^{-1}$,
of the sum of the squares of the first $N+1$ F-B. functions.
Gaussian averaging is not as mandatory here as it is in the study
of individual F-B. functions, since its regulating r\^ole can be
also supplied by the summation over $n$, and yet I am not aware of
any rigorous treatment involving only the Fourier kernel
$\chi_{\omega}(r-s)=e^{-i(r-s)/\omega}$. In any case, we shall
maintain this ambiguity offering theoretical results that require
averaging, and---at times---experimental results showing that
averaging can be disposed of.

In physical language, the discrete probability distribution
$|\psi_n(t)|^2=|{\cal J} _n(\mu;t)|^2$ (recall the normalization
condition Eq. (\ref{norma})) is called the {\em wave--packet}, and
therefore $\nu_0(N,\omega)$ is the {\em distribution function} of
the Gaussian averaged wave--packet. It therefore contains all the
information on this probability distribution, and a detailed
control of this quantity, in $N$ and $\omega$, extends to the
growth exponents.

Typically, {\em upper} bounds on $\nu_0(N,\omega)$ have been
found, yielding {\em lower} bounds on growth exponents for
positive $\alpha$. This can be easily seen by remembering that the
quantum probability distribution $|\psi_n(t)|^2$ is normalized 
by eq. (\ref{norma}):
$\lim_{N \rightarrow \infty} \nu_0(N,\omega) = 1$ for all
$\omega$; therefore, squeezing the head of the distribution fatten
its tail. The original result is Guarneri's inequality
$\beta^-(\alpha) \geq D_1(\mu)$, extended by Combes \cite{combes}
to many-dimensional Schr\"odinger operators, further refined by
Guarneri and Schulz-Baldes \cite{gsb2}, and by Tcheremchantsev et
al. \cite{tre2,tcl} to a moment-dependent bound, in the form
 \beq
   \beta^-(\alpha) \geq D_{(1+\alpha)^{-1}}(\mu).
   \nuq{lowint}
In the above, $D_q(\mu)$ are the generalized dimensions of the
measure $\mu$, of index $q$, that we shall define in Sect.
\ref{sgend}. The original hypothesis \cite{gsb2} that these
dimensions exist for all $q \in {\bf R}$, and are finite for some
$q<1$ has been weakened \cite{tcl} to cover the case of the most
general positive Borel measure $\mu$. Notice finally that these
bounds involve generalized dimensions of {\em positive} index,
between zero and one. Inspection of the proofs reveals that this
is a limitation of the technique, that deals rather crudely with
the role of the orthogonal polynomials $p_n(\mu;x)$.

\section{Further lower bounds to the growth exponents}
\label{slow}

\begin{figure}[ht]
\includegraphics[height=10cm,width=14cm]{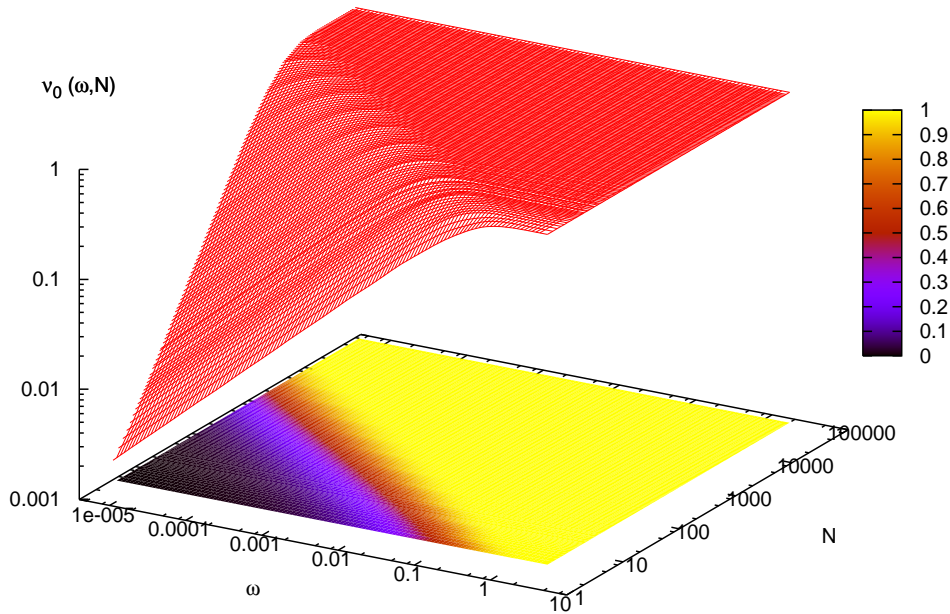}
\caption{Truncated, averaged moment $\nu_0(N,\omega)$ for a Julia
set measure with $\lambda = 2.9$. To the left of the figure, the
region where the ansatz (\ref{ketz1}) is well verified. The flat
plateau to the right, at $\nu_0(N,\omega)=1$, stretches over all
values of $N$ that at time $t=\omega^{-1}$ have not yet been
reached by the propagating wave. } \label{fbes3}
\end{figure}

An improvement of these estimates is obtained if one controls the
growth rate of orthogonal polynomials. The first attempt in this
direction has been the renormalization theory of orthogonal
polynomials of IFS measures \cite{physd1,physd2} that we shall
meet in the following. Successively, the imaginative formula for
the function $\nu_0(N,\omega)$ proposed by Ketzmerick et al.
\cite{ket} opened a different perspective:
 \beq
\nu_0(N,\omega)\sim N^{d}\omega^{D_{2}(\mu)},
 \nuq{ketz1}
where $D_{2}(\mu)$ is the correlation dimension of the measure
$\mu$ (see Sect. \ref{sgend}) and $d$ is a suitable constant that
depends on properties of the orthogonal polynomials of $\mu$, as
we shall discuss momentarily \cite{notad2ps}.
Formula (\ref{ketz1}) can obviously be valid only for $1 \ll N \ll
\omega^{-D_{2}(\mu)/d}$. It predicts a scaling form, both in time
and space, of the wave--packet. Pictorially, the authors of
\cite{ket} say that, at fixed time, the initial part of the
wave-packet decays with $n$ as $n^{d-1}$. In Figure \ref{fbes3}
the function $\nu_0(N,\omega)$ is plotted versus $\omega$ and $N$,
in the case of Figs. \ref{fbes1} and \ref{wave4}. We observe that
the scaling (\ref{ketz1}) is well verified, in the region in
space--time that corresponds to the decay of excitations, behind
the wave--front.

Starting from formula (\ref{ketz1}) Ketzmerick et al. have derived
a lower bound to the growth exponents in the form $\beta(\alpha)
\geq D_2(\mu)/d$. This result can be put on rigorous footing
recalling the observation that the square moduli of all F-B.
functions decay as $t^{-D_2(\mu)}$ \cite{pagen}. Therefore
 $\nu_{0}(N,\omega) \,\omega^{-D_2(\mu)}
 $
 is a bounded function of $\omega$. If in addition there exist
 $\gamma$ and $c$ larger than zero such that
\beq
  \nu_0(N,\omega) \leq  c N^{\gamma}\omega^{D_{2}(\mu)},
 \nuq{gal272}
for all $N$ and $\omega$, we can conclude that
   \beq
 \beta_-(\alpha)\,\geq\,\frac{D_2(\mu)}{\gamma}\;,
 \label{gal14}
\eeq
 for all positive values of $\alpha$.
In Sect. \ref{sjulia1} we shall comment on the effectiveness of
this bound in an exactly computable situation. Notice that our
definition of $\gamma$ over-estimates the parameter $d$ in
Ketzmerick et al. surmise (\ref{ketz1}), and eq. (\ref{gal272})
may be too crude of an estimate. Yet, in certain cases, numerical
experiments as that of Fig. \ref{fbes3} show that in a region of
space--time the surmise is a good description of the function
$\nu_0(N,\omega)$, and $d$ is a good approximation of $\gamma$. As
a matter of facts, the exponent $d$ has a dimensional flavor,
which mixes the asymptotic properties of the orthogonal
polynomials $p_n(\mu;s)$ and the local properties of the measure
$\mu$. To see this, it is now time to briefly introduce the
generalized dimensions of a measure $\mu$.

\section{Generalized dimensions of the orthogonality measure}
\label{sgend}
The spectrum of generalized dimensions $D_q(\mu)$ of a positive
measure $\mu$ is given, for real $q \neq 1$, by the law
 \beq
 \int\!d\mu(r)  (\mu(B_\omega(r)))^{q-1} \sim
 \omega^{(q-1)D_q(\mu)}.
 \label{term3}
 \eeq
The scaling law is made precise by taking superior and inferior
limits, when $\omega$ tends to zero, of the logarithm of the
l.h.s. integral over the logarithm of $\omega$. Of course, an
appropriate formula exists also for $q=1$. A thorough study of
generalized dimensions is to be found in \cite{pesin}, \cite{tre}.

We mention now for future reference an alternative approach to the
evaluation of the scaling law (\ref{term3}). Think of covering the
support of $\mu$ by a family $\Sigma$ of disjoint intervals
$I_\sigma$, of length $l_\sigma$, and measure $\pi_\sigma :=
\mu(I_\sigma)$. Then, $D_q(\mu)$ is defined as the divergence
abscissa of $H(x,\Sigma)$,
 \beq
% \lim_{k\,\rightarrow\,\infty}
 H(x,\Sigma) := \sum_{\sigma \in \Sigma }
 \pi_\sigma^{q} \;l_{\sigma}^{(1-q)x},
 \label{ev452b}
\eeq when the generalized limit of finer and finer coverings is
taken.

We can now understand  why the correlation dimension have a r\^ole
in our problem \cite{theo}. Let us start from the expansion
 \beq
 {\cal A}_G(|{\cal J}_n(\mu;t)|^2) =
 \int\!\!\int d\mu(s)d\mu(r) \chi_{\omega}(r-s)
 p_{n}(\mu;s)p_{n}(\mu;r),\;
\label{term1}
 \eeq
and observe that, when $\omega \ll 1/n$, the variation of
$p_{n}(\mu;s)$ over $B_\omega(r)$, the ball of radius $\omega$
centered at $r$, is negligible, so that $p_{n}(\mu;s) \simeq
p_{n}(\mu;r)$ and
 \beq
  {\cal A}_G(|{\cal J}_n(\mu;t)|^2)
  \;\;\;
   \simeq
  \;\;\;
\int\!d\mu(r) p_{n}^2(\mu;r) \!\int d\mu(s)\chi_{\omega}(r-s)
 \;\;\;
\simeq \;\;\;
 \int\!d\mu(r) p_{n}^2(\mu;r) \mu(B_\omega(r)).
\label{term2}
 \eeq
Now, the correlation dimension $D_2(\mu)$ is obtained setting
$q=2$ in eq. (\ref{term3}). It so happens that the function
$p_{n}^2(\mu;r)$ does not alter the asymptotic behavior of the
last integral in eq. (\ref{term2}), and therefore $D_2(\mu)$
governs the asymptotic decay of the averaged square moduli of F-B.
functions. Of course, this is not a substitute for a rigorous
proof, that has been obtained in a variety of ways in the
literature, as explained in detail in \cite{pagen}.

\section{Asymptotics of the orthogonal polynomials and growth exponents}
\label{sgend2}

We can now return to the wave--propagation problem, and  apply the
same approximation as in eq. (\ref{term2}) to $\nu_0(N,\omega)$,
to get
 \beq
  \nu_0(N,\omega)
\simeq \int\!d\mu(r)  \mu(B_\omega(r))
  \sum_{n=0}^N  p_{n}^2(\mu;r) .
\label{term4}
 \eeq
Suppose now that the orthogonal polynomials verify a scaling
relation of the kind
 \beq
 \sum_{n=0}^N  p_{n}^2(\mu;r) \sim g(r) N^{d(r)},
 \nuq{term5}
 for large $N$, with local dimension $d(r)$,  and a smooth
function $g(r)$ (where smooth is intended as a subleading
behavior).
 Then one meets the problem, familiar in dimension theory, of
determining the exponent $d(\omega)$, defined by
 \beq
 \int\! d\varrho_\omega (r) g(r) N^{d(r)} \sim N^{d(\omega)},
\label{term5b}
 \eeq
in terms of the measures $d \varrho_\omega (r) := \mu(B_\omega(r))
d\mu(r)$. Suppose now that there exists constants $C$ and $\gamma$
such that
 \beq
   \sum_{n=0}^N p^2_n(\mu;x) \leq C N^\gamma
 \nuq{lgamma1}
 for any $x$ in the support of $\mu$,
then Ketzmerick {\em et al.} surmise holds. A more refined
analysis \cite{gsb2,tcl} can be carried on restricting the
integral with respect to $\mu$ to appropriate subsets $\Omega$ of
the support of $\mu$, so to obtain lower bounds to
$1-\nu_0(N,\omega)$. This analysis shows that indeed under the
above hypothesis \ref{lgamma1} the following lower bound holds:
  \beq
   \beta^-(\alpha) \geq \frac{1}{\gamma}
   D_{(1-\alpha/\gamma)}(\mu).
   \nuq{lowint2}
Two comments are in order: the first, is that eq. (\ref{lowint2})
is a better bound than $D_2(\mu)/\gamma$. The second, that
generalized dimensions of argument less than one appear. We shall
soon return to this fact.

 In the same line is the local result
\cite{last}: suppose that there exists a Borel set $S$ of positive
measure, so that the restriction of $\mu$ to this set is
$a$-continuous (it gives zero weight to any set of null
$a$-dimensional Hausdorff measure) and so that there exists
$\gamma$ such that for any $x \in S$ (\ref{lgamma1}) is verified,
then
 \beq
   \beta^-(\alpha) \geq \frac{a}{\gamma}.
   \nuq{lowint3}
Further lower bounds are described in \cite{tre3}, \cite{tcl}.

\section{Upper bounds to the quantum intermittency function}

Lower bounds on $\nu_0(N,\omega)$ do not lead to upper bounds on
$\beta(\alpha)$ \cite{last}, that are therefore much harder to
find \cite{physd2,itaupp,hermann,combgio}, also because the
strategy of restricting the consideration to a subset $\Omega$ of
the support of $\mu$ is not sufficient here.

The last quoted reference describes a rather interesting situation
that is worth presenting in some detail, also because the
techniques on which it is based might find wider applicability.
One starts from the Jacobi matrices ${\bf J}_{\theta,\eta}$
introduced in \cite{jl1} and defined by
 \beq
  x p_k(x;\mu) = (V_\eta(k)+
  \theta \delta_{0,k} ) p_k(x;\mu)
 + p_{k-1} (x;\mu) + p_{k+1} (x;\mu)
 \nuq{jl1}
labelled by the real parameters $\theta \in [\theta_0 ,
\theta_1]$, $0 < \theta_0 < \theta_1 < \infty$ and $\eta \in
(0,\infty)$. The structure of the recurrence relation renders
${\bf J}_{\theta,\eta}$ a discrete Schr\"odinger operator, with
potential $V_\eta(k)$.  This is chosen to be null, except on a set
$B$ of selected {\em barrier} locations, $B = \{ L_n, n \in {\bf
Z}^+ \}$:
 $
   V_\eta(k) = \chi_B(k) \; k^{\eta}.
 $
The exponent $\eta$ links location and height of the barrier. The
limit $\eta=\infty$ corresponds to a Dirichlet condition at each
$L_n$, that clearly means no propagation and pure point spectrum.
On the other hand, $\eta=0$ gives barriers of constant height, and
generically absolutely continuous spectrum if these are sparse
enough. Sparseness is a convenient request for analysis: assume
that for some $a > 1$ and all $n \in
 {\bf Z}$
  $
   L_{n+1} \geq a L_n.
  $
Under these conditions one can prove \cite{combgio} that for all
$\alpha$, $0 < \alpha \leq 2$, and almost all
 $\theta$ one has that
  \beq
   \beta^-(\alpha) \leq \frac{\alpha + 1}{2 \eta + \alpha + 1},
  \nuq{bound3}
while $\beta^+(\alpha) = 1$: there is a part of the wave-packet
that moves linearly in time (one says ballistically, in the usual
jargon) while the main body follows at a slower pace. We refer to
\cite{combgio} for the detailed analysis and illustrative
pictures.

\section{Wave--front propagation: unsolved asymptotics}
\label{tailsec} The previous sections have dealt with the shape of
the wave--packet {\em behind} the wave--front. This implies that
time, the argument of the F-B. functions, is much larger than
space, the order. To complete the picture we must take into
account what happens in the opposite limit, and, more importantly
for our goals, in the region of the wave--front. This will explain
our remark of Sect. \ref{sshape}, on the fact that lower bounds on
$\nu_0(N,\omega)$ in the first region do not yield control of
$\beta(\alpha)$.

\begin{figure}[ht]
\includegraphics[height=10cm,width=14cm]{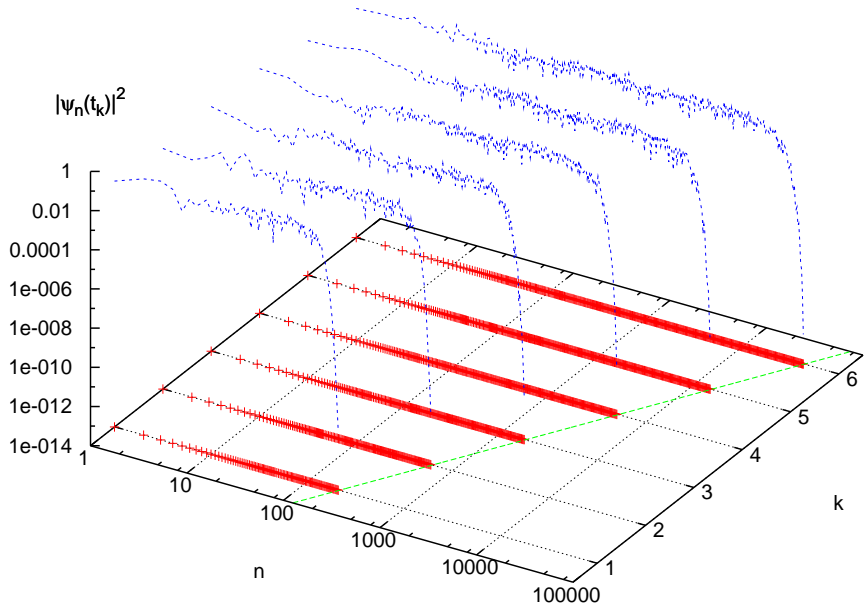}
\caption{Snapshots of $|\psi_n(t_k)|^2$ at exponentially spaced
times $t_k$, $k=1,\ldots,6$, versus $N$ and $k$ for an I.F.S.
measure, described in Sect. \ref{ifssec}. At the bottom of the
graph, the fitting line for the wave--front indicates the law $n
\sim t^{\eta}$, with $\eta=0.84165$.} \label{figwave}
\end{figure}

A sequence of snapshots illustrating the wave-packet at
exponentially spaced times  is shown in Fig. \ref{figwave}. Over
the time--span of the figure, the wave enlarges its size by more
than two orders of magnitude. Two characteristics are to be
remarked: the decay of the wave--packet is clearly consistent with
eq. (\ref{tev2}), but in addition it takes place rather abruptly
{\em past} a wave--front position. The motion of this point, on
the other hand, appears in Fig. \ref{figwave} to follow an
algebraic law, with exponent $\eta$. These two characteristics
combined imply an {\em upper} bound to the growth exponents,
$\beta^+(\alpha) \leq \eta$, for all positive values of $\alpha$.

In the usual Bessel case, it is well known that $\eta=1$ (Bessel
functions decay abruptly when the order exceeds the exponent), and
that the wave is propagating linearly, so that a speed of
propagation can be defined. It is interesting to remark that
Newton's determination of the speed of sound ultimately relies on
these properties \cite{bril},\cite{physd2}. In the singular
measure case, a speed {\em cannot} be defined, and we are forced
to introduce the intermittency function $\beta$ of the moment
order $\alpha$. In view of this, and of our previous remark on the
difficulty of obtaining upper bounds to the growth exponents, it
is of crucial relevance to develop techniques to control this
particular asymptotics of the F-B. functions, so to enable us, for
instance, to characterize the exponent $\eta$ observed in Fig.
\ref{figwave}.

\section{Julia Set Measures: renormalization equations}
\label{sjulia1}
 We now discuss an example that can be worked out
exactly to a large extent. We choose for $\mu$ the balanced
measure supported on a real Julia set, generated by the quadratic
map $z \rightarrow z^2 -\lambda$, for $\lambda \geq 2$
\cite{barn,danbel}. The inverses of this map,
 \beq
 \phi_j (s) = j \; \sqrt{s+\lambda}
 \label{ev427x}
 \eeq
 % and probabilities $\pi_\sigma = \frac{1}{2}$.
 with $j =  \pm 1$,
can be seen as the non-linear maps of an Iterated Function System.
The invariant measure of this I.F.S. is defined via the equation
 \beq
 \int f(x) d \mu(x) = \frac{1}{2} \sum_{j = \pm 1}
  \int (f \circ \phi_j) (x) d \mu(x),
  \nuq{balabiot}
valid for any continuous function $f$. When $\lambda=2$, we obtain
the orthogonality measure of the Chebyshev polynomials, suitably
rescaled. When $\lambda
> 2$, the support of $\mu$ is a real Cantor set. In the graphs
displayed in this paper, we have chosen $\lambda=2.9$ for no
particular reason.

The hierarchical structure of the support of $\mu$ is brought to
evidence by iterating the I.F.S. maps $k$ times: to keep the
notation compact it is useful to define the index vector ${\bf
\sigma}=(\sigma_1,\ldots,\sigma_k)$, with $\sigma_i \in
\{+1,-1\}$,  of length $|\sigma|=k$, and the associated
composition maps $\phi_\sigma=\phi_{\sigma_1} \circ
\phi_{\sigma_2} \circ \cdots \circ \phi_{\sigma_k}$. Let now
$I_\emptyset$ be the convex hull of the support of the measure
$\mu$, $I_\emptyset = [-\Lambda,\Lambda]$, where $\Lambda$ is the
fixed point of $\phi_+$. At hierarchical order $k=|\sigma|$, the
support of $\mu$ is covered by the intervals $I_\sigma$,
 \beq
   I_\sigma := \phi_\sigma (I_\emptyset).
  \nuq{isigma}

The following remarkable property holds for the orthogonal
polynomials of this measure \cite{ptic,barn}:
 \beq
 p_{2n} (\mu;\phi_j (s)) = p_n(\mu;s),
 \nuq{balab2}
for $j = \pm 1$. Applying this property, and  the balance equation
(\ref{balabiot}) to the Gaussian time averaged wave-function
projections, $ {\cal A}_G(|{\cal J}_n(\mu;t)|^2)$, that we denote
for short $\psi^G_{n}(t)$, we get:
 \beq
  \psi^G_{2n}(t) =
 \frac{1}{4}
 \sum_{\sigma,\sigma'}
  \int\!\!\int
     d\mu(s)\,d\mu(r) \;
 \chi_{\omega} (\phi_\sigma(r)-\phi_{\sigma'}(s))
  p_n(\mu;r) p_n(\mu;s).
  \nuq{reno1}
Iterating the renormalization procedure $k$ times, we obtain the
wave-function average projection at site $N=n 2^k$, with $n$ and
$k$ integers, in the form:
 \beq
  \psi^G_{N}(t) = \frac{1}{2^{2k}}\!\!\!\!\!\!\!\!
 \sum_{\scriptsize \ba{c}  \sigma,\sigma' \\
  % \mbox{\em \scriptsize  s.t.  }
  |\sigma|=|\sigma'|=k
  \ea }
    \!\!\!\!\!\! \!\!
  \int\!\!\int
     d\mu(s)\,d\mu(r) \;
   \chi_\omega(\phi_\sigma(r)-\phi_{\sigma'}(s)) \;
 p_n(\mu;r) p_n(\mu;s).
   \nuq{bals2b}

The non-diagonal contributions, $\sigma \neq \sigma'$, in the
balance equation (\ref{bals2b}) have a fast time decay and can be
neglected. In addition, in the diagonal terms, the non-linear maps
$\phi_\sigma$ can be replaced by a linearized version: for any
$\sigma$, let now
 \beq
  l_\sigma(s) :=  \delta_\sigma s + \theta_\sigma
  \nuq{lifs1}
  be the linear map,
with coefficients $\delta_\sigma$ and $\theta_\sigma$, that takes
$I_\emptyset $, the convex hull of the support of the measure
$\mu$, exactly onto $I_\sigma$. In other words, $I_\sigma =
\phi_\sigma(I_\emptyset) = l_\sigma(I_\emptyset)$, and the length
of this cylinder is consequently proportional to $\delta_\sigma$:
$|I_\sigma| = 2 \delta_\sigma  \Lambda$. Usage of linear maps in
the argument of $\chi_\omega$ has the effect of dividing $\omega$
by $\delta_\sigma$, so that
 \beq
 \psi^G_{N}(t) = \frac{1}{2^{2k}}
 \sum_{\sigma \mbox{\em \scriptsize  s.t.  } |\sigma|=k}
 \psi^G_{n}(t \delta_\sigma) +
  {\cal E} (k,n,t),
  \nuq{lereno}
where ${\cal E} (k,n,t)$ is the error involved in the
approximations we have made. The related error estimates are
rather involved, and aim to show that ${\cal E} (k,n,t)$ is
negligible, in appropriate asymptotic expansions. We shall boldly
do this in the following.

Eq. (\ref{lereno}) is a renormalization equation, that links the
wave-function projection at site $N$ and time $t$ to those at site
$n$ and earlier times. As opposed to simple estimates of
$\nu_0(\omega,N)$, this equation offers us a means of controlling
the growth exponents. We have developed this idea in
\cite{physd1,physd2} and in the more rigorous, yet less noticed
ref. \cite{france}.

\section{Julia set measures: analysis behind the wave front}
We now employ the renormalization analysis, eq. (\ref{lereno}), to
compute exactly the behavior of $\nu_0(N,\omega)$ in the region
behind the wave front:
 \beq
  \nu_0(N,\omega) \sim \sum_{j=0}^{n-1} 2^k \psi^G_{j2^k}(\omega^{-1})
  \sim
  % \!\!\!\!\!\! \!\!
 \sum_{\sigma \mbox{\em \scriptsize  s.t.  } |\sigma|=k}
 \sum_{j=0}^{n-1}
 \frac{1}{2^{k}} \psi^G_{j}(\delta_\sigma \omega^{-1}) =
 % \!\!\!\!\!\! \!\!
  \sum_{\sigma \mbox{\em \scriptsize  s.t.  }
 |\sigma|=k} \frac{1}{2^{k}}
 \nu_0(n,\omega/\delta_\sigma).
 \nuq{leren2}
 The form $\nu_0(N,\omega)\,\sim\,N^{\gamma}\;\omega^{\,D_{2}(\mu)}$
solves eq. (\ref{leren2}); upon setting $\pi_\sigma := 2^{-k}$, we
get
 \beq
  \sum_{\sigma \mbox{\em \scriptsize  s.t.  }
 |\sigma|=k}
 \pi_\sigma^{1+\gamma} \;\delta_{\sigma}^{-D_{2}(\mu)}  \sim 1,
 \label{ev452}
\eeq
  that implicitly determines $\gamma$. This determination
  is indeed transparent:
comparing eq. (\ref{ev452}) with the discrete evaluation of the
generalized dimensions, eq. (\ref{ev452b}), we immediately obtain
  \beq
 D_{2}(\mu)\;=\;\gamma\;D_{1+\gamma}(\mu)\;,
 \label{ev453}
\eeq
  and therefore
\beq
 \gamma = 1.
 \label{ev454}
\eeq

\begin{figure}[ht]
\includegraphics[height=10cm,width=14cm]{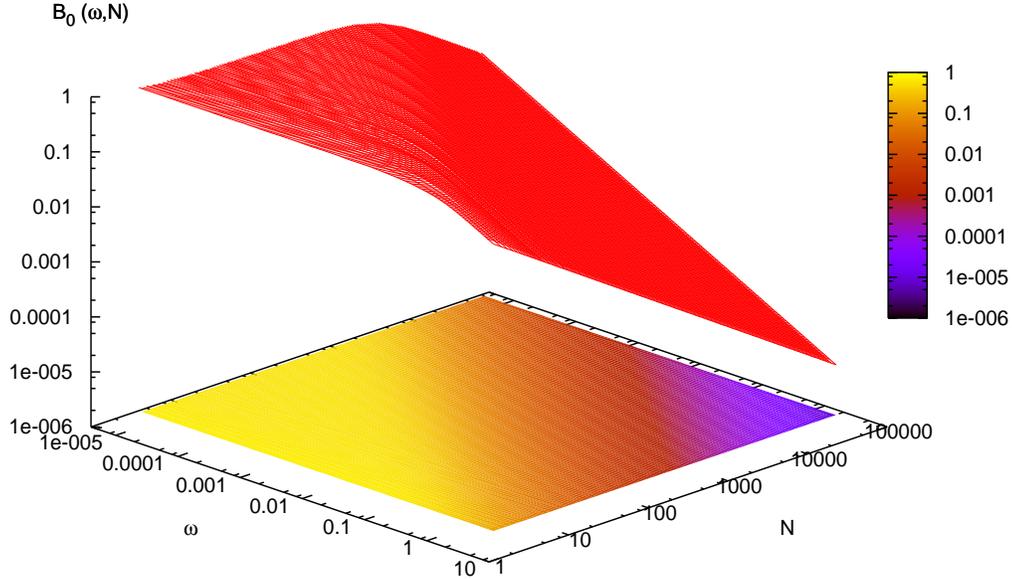}
\caption{$B_0(\omega,N):=N^{-1} \omega^{-D_2(\mu)}\nu_0(N,\omega)$
for the Julia set measure of Fig. \ref{fbes3}.} \label{fbes4}
\end{figure}

Figure \ref{fbes4} shows the function $B_0(\omega,N):=N^{-1}
\omega^{-D_2(\mu)}\nu_0(N,\omega)$, whose flat left piece confirms
the validity of the scaling (\ref{ketz1}), and of the value
$\gamma=1$.

According to Sect. \ref{slow}, a consequence of this calculation
is a lower bound on the positive exponents: $\beta(\alpha) \geq
D_2(\mu)$. Notice that since $D_2(\mu) < D_1(\mu)$ this bound is
weaker that the original Guarneri's inequality $\beta(\alpha) \geq
D_1(\mu)$ \cite{notabd}. This fact is by no means accidental:
information of the kind (\ref{ketz1}), and more general, on the
$\nu_0(N,\omega)$ for small $N$ (with respect to an appropriate
power of $\omega$) is not sufficient to control the growth
exponents.

\section{Surfing the Intermittent Quantum Wave}
\label{ssurf}
 A treatment quite analogous to that of the previous
section can be carried out for all truncated moments of order
$\alpha >0$:
 \beq
   \nu_\alpha(N,\omega):=
   \sum_{n=0}^{N} n^\alpha {\cal A}_G(|{\cal J}_n(\mu;t)|^2) =
  \int\!\!\int
     d\mu(s)\,d\mu(r)
     \chi_{\omega}(r-s)
%   e^{-\frac{(r-s)^2}{\omega^2}}
   \sum_{n=0}^{N}\,  n^\alpha p_{n}(\mu;s)\,p_{n}(\mu;r)\;.
   \label{gal4a}
\eeq
\begin{figure}[ht]
\includegraphics[width=14cm,height=10cm]{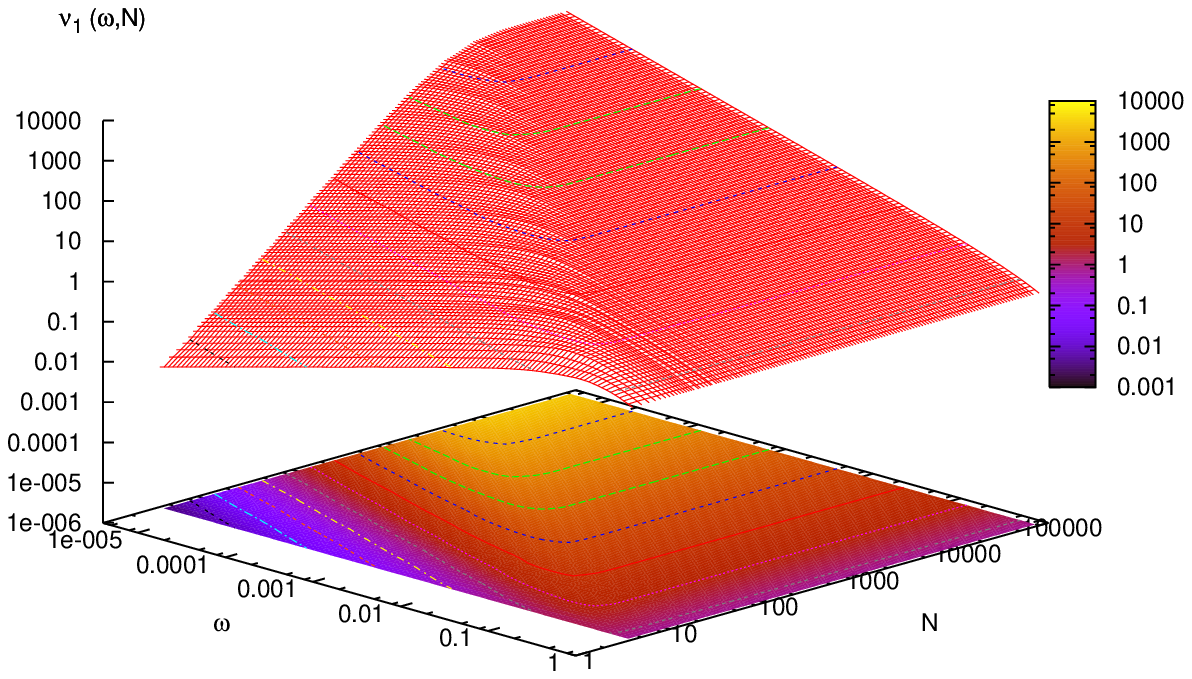}
\caption{Gaussian averaged, truncated first moment
$\nu_1(\omega,N)$ for the Julia set measure with $\lambda = 2.9$.}
\label{fjulal}
\end{figure}
\begin{figure}[ht]
\includegraphics[width=14cm,height=10cm]{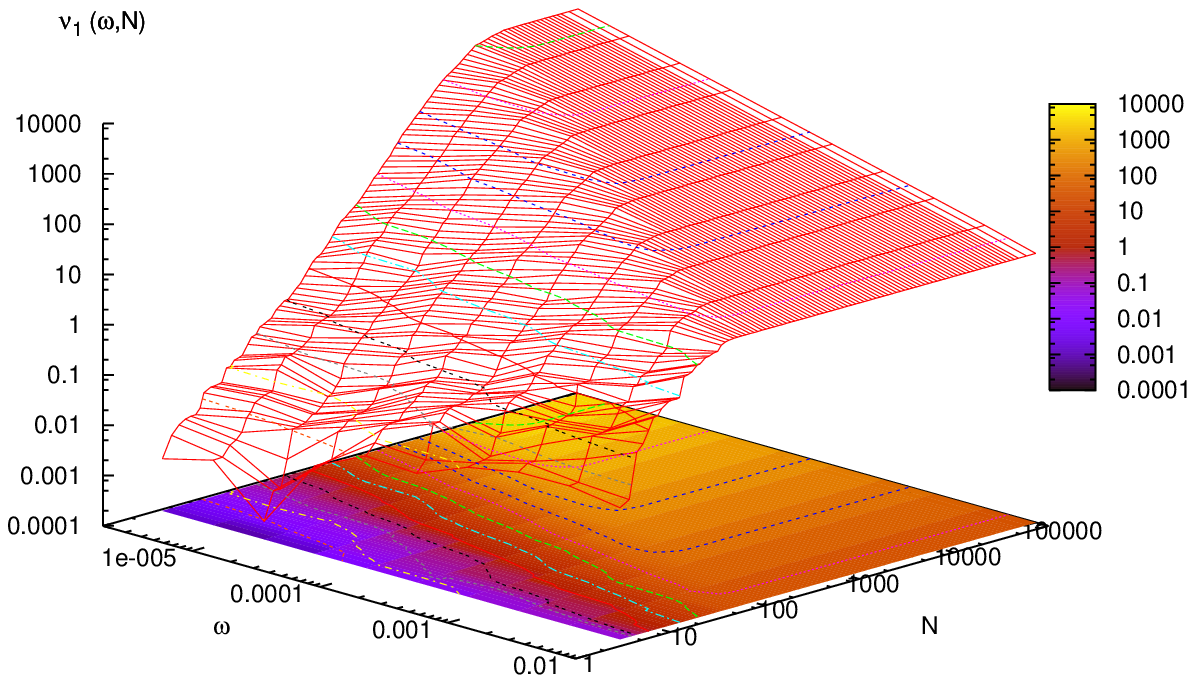}
\caption{Truncated first moment $\nu_1(\omega,N)$ for the Julia
set measure with $\lambda = 2.9$ {\em without} time--averaging.
Notice the decay of fluctuations with increasing values of $N$.}
\label{fjula2}
\end{figure}
Using the renormalization eq. (\ref{lereno}) in the new situation
leads to the result
 \beq
   \nu_\alpha(N,\omega)\,\sim\,N^{\gamma+\alpha}\;\omega^{\,D_{2}(\mu)},
\label{ev457}
 \eeq
 valid in the regime of decaying F-B. functions, in the leftmost part of
Fig. \ref{fjulal}, where this behavior is clearly observed.

Notice that a new scaling region appears now to the right of the
figure, {\em ahead} of the wave front, replacing the plateau that
was obtained for $\alpha=0$: in fact, when the lattice site $N$
has not yet been reached by the wave, $\nu_\alpha(N,\omega)$ is
independent of $N$, and is equal to the (Gaussian averaged)
position moment of order $\alpha$. Therefore, in such region,
 \beq
   \nu_\alpha(N,\omega)\,\sim\,N^{0}\;\omega^{-\alpha \beta(\alpha)},
\label{ev458} \eeq
 in which the intermittency function $\beta(\alpha)$ explicitly
appears. Of course, this is a trivial observation. It can be
turned into a constructive theory only if we can stretch our
approximations to reach this region.

 Pictorially, but
appropriately, we can say that the lower bounds mentioned in the
previous sections have been obtained by floating safely in the
calm waters behind the wave--front. To the opposite, a complete
theory of quantum intermittency can be obtained only if we are
brave enough to {\em catch the wave}, boldly surfing on our
approximation board the roaming waters of the F-B. wave--front,
vividly depicted in Figure \ref{wave4}.

Achieving this goal is a rare accomplishment: the renormalization
approach is the board that has enabled us to do this for Julia set
measures \cite{physd1,physd2,france}. First,
 \beq
 \nu_\alpha(t) \sim \sum_{j=0}^{\infty} 2^k (j 2^k)^\alpha
  \psi^G_{j2^k}(t).
  \nuq{pis1}
Then, employing again the renormalization eq. (\ref{bals2b}), we
obtain
 \beq
 \nu_\alpha(t) \sim
\sum_{\sigma \mbox{\em \scriptsize  s.t.  } |\sigma|=k}
 {2^{k(\alpha-1)}}
 \sum_{j=0}^{\infty}
  j^\alpha \psi^G_{j}(\delta_\sigma t) =
  {2^{k(\alpha-1)}}
 \!\!\!\!\!\! \!\!
  \sum_{\sigma \mbox{\em \scriptsize  s.t.  }
 |\sigma|=k}
\!\!\!\!\!\! \!\!
 \nu_\alpha(\delta_\sigma t)
  \nuq{pis2}
This relation has the scaling solution $\nu_\alpha(t) \sim
t^{\alpha \beta(\alpha)}$, whence by consistency
 \beq
  \sum_{\sigma \mbox{\em \scriptsize  s.t.  }
 |\sigma|=k}
 \pi_\sigma^{1-\alpha} \;\delta_{\sigma}^{\alpha \beta(\alpha)}  \sim 1,
 \label{ev452c}
\eeq
 that unveils, by comparison with eq. (\ref{ev452b}),
the fundamental Julia set relation
  \beq
 \beta(\alpha)\;=\;D_{1-\alpha}(\mu)\;,
 \label{jul10}
\eeq
  that links growth exponents and generalized dimensions.

We have verified numerically that this relation holds exactly even
{\em without} time--averaging \cite{physd1,physd2,france}. Indeed,
Fig. \ref{fjula2} displays the truncated, instantaneous value of
the first moment versus $\omega$ and $N$: as we have remarked
previously, summation over $n$ supplies the regularizing effect.

In \cite{piech} a relation formally written as eq. (\ref{jul10}),
but different in meaning, has been obtained by a renormalization
procedure over Fibonacci Jacobi matrices. In such relation
$\beta(\alpha)$ are the growth exponents of moments {\em averaged
over initial sites} (which mathematically amounts to averaging
over different Jacobi Hamiltonians), and $D_q$ are the
thermodynamical dimensions of the logarithmic potential
equilibrium measure, that we shall also consider in the next
section. A proof of this result for a family of Jacobi matrices
has been obtained recently \cite{bellitalo}.

\section{Linear I.F.S.: renormalization theory and a conjecture}
\label{ifssec} The results obtained in the previous three sections
are certainly neat, but by no means universal. They stem from the
clean renormalization properties of Julia set orthogonal
polynomials, eq. (\ref{balab2}), in a situation characterized by
other remarkable symmetries, the most notable of which is perhaps
the fact that the measure of the zeros of $p_n(\mu;s)$, the
logarithmic potential equilibrium measure, coincides with $\mu$
itself. In addition, it is clear that we cannot approximate an
arbitrary measure with Julia set measures.

\begin{figure}[ht]
\includegraphics[width=9cm,height=12cm,angle=270]{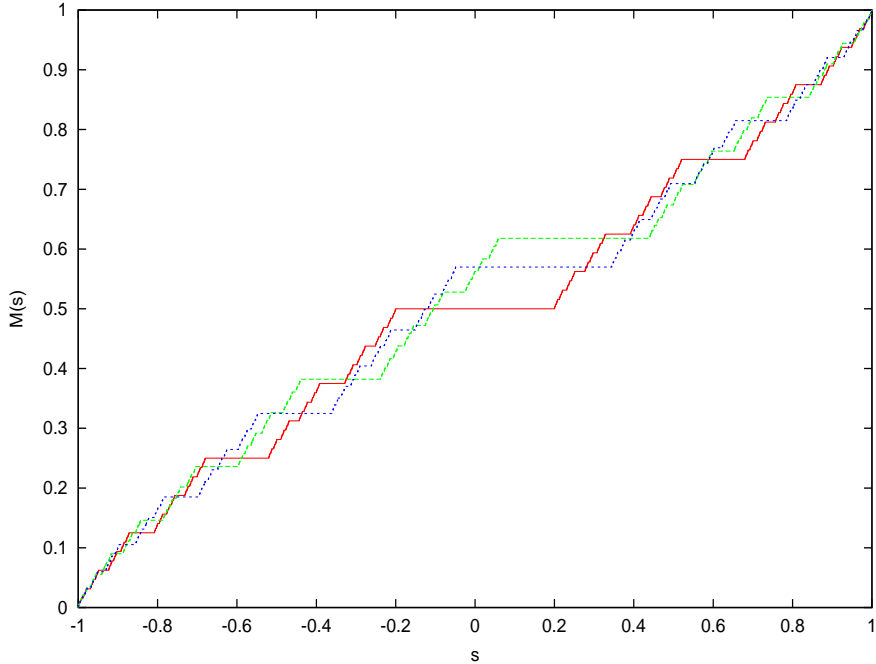}
\caption{Distribution functions $M(s):=\int_{-\infty}^s d\mu(s')$
for three uniform Gibbs measures with the same $D_q(\mu)=D_0={\log
2}/({\log 5 - \log 2})$, as described in the text.} \label{distr}
\end{figure}

To the contrary, linear iterated function systems
\cite{hut,dem,ba2,gohw}, in which we have at our disposal an
unlimited number of maps of the kind (\ref{lifs1}), $l_i(s)
=\delta_i s + \theta_i$, and associated probabilities $\pi_i$,
$i=1,\ldots,M$ can approximate arbitrarily well any measure with
bounded support \cite{giohandy}. These I.F.S. define invariant
measures $\mu$ via the obvious generalization of eq.
(\ref{balabiot}),
  \beq
 \int f(s) d \mu(s) =   \sum_{i=1}^M
 \pi_i  \int (f \circ l_i) (s) d \mu(s).
  \nuq{balabiot2}

Moreover, linearity of the maps implies the renormalization
equation
\begin{equation}
   p_{n} (\mu;l_i(s)) =
                \sum_{k=0}^{n} \Gamma_{i,k}^{n} p_k(\mu;s).
\label{est1}
\end{equation}
The coefficients $\Gamma$ have a profound meaning, as they are the
{\em Lanczos vectors}  associated with a generalization of the
Jacobi matrix ${\bf J}_\mu$ \cite{cap}. In ref. \cite{physd2} I
have employed eqs. (\ref{balabiot2}),(\ref{est1}) in a similar
fashion than in Sect. \ref{ssurf}, to study the intermittency
function $\beta(\alpha)$. It has not been possible, though, to
close the asymptotic relations exactly, but only to obtain a
sequence of approximation of the intermittency function.
Nonetheless, this approximate renormalization theory has shown
that the key to the asymptotic behavior of the moments
$\nu_\alpha(t)$ lies in the properties of the coefficients
$\Gamma$. Needless to say, these properties are rather elusive.

\begin{figure}[ht]
\includegraphics[width=12cm,height=9cm]{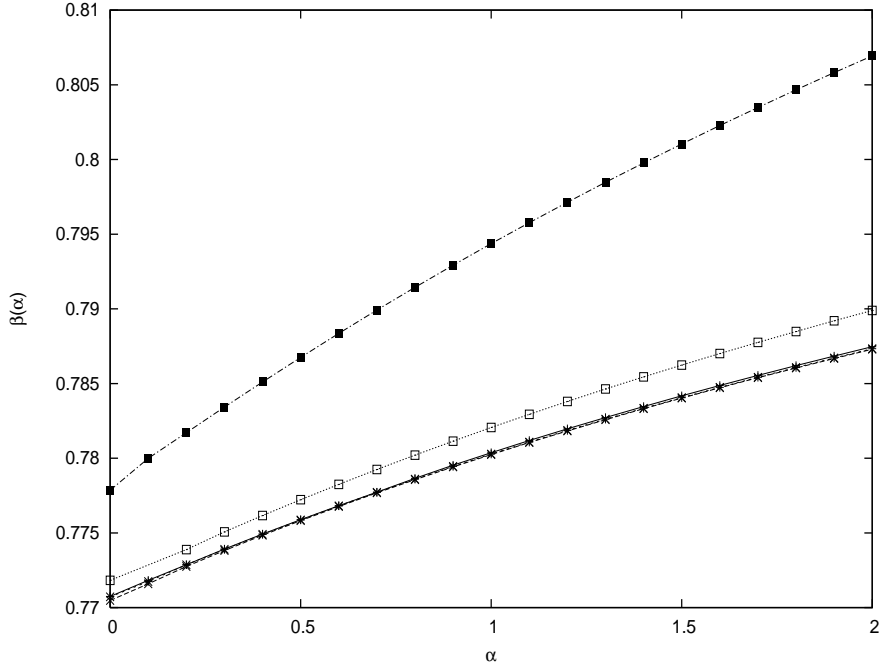}
\caption{Intermittency function $\beta(\alpha)$ for I.F.S.
generated uniform Gibbs measures with the same spectrum of
generalized dimensions, $D_q(\mu)=D_0={\log 2}/({\log 5 - \log
2})$. Data plotted are for: two--maps I.F.S. (three lower
coincident curves, crosses); a symmetrical three--maps I.F.S.
(central curve, open squares), and an asymmetrical three--maps
I.F.S. measure (top curve, filled squares).} \label{zorro}
\end{figure}

In closing this paper I want to discuss an additional piece of
evidence from \cite{physd1} that might give us a clue on the
general problem, and is still (to my knowledge) unexplained.
Consider the set of linear I.F.S. generated by just two maps, for
which
\begin{equation}
\label{gold1}
  \pi_i = \delta_i^{D_0}, \;\; i=1,2
\end{equation}
where $D_0$ is a real number between zero and one, that must
obviously satisfy the probability conservation equation
 \beq
   1 = \sum_{i=1}^2 \pi_i  = \sum_{i=1}^2 \delta_i^{D_0}.
   \nuq{gibbs0}
Because of the latter equality, $D_0$  is the box-counting
dimension of the support of $\mu$. Clearly, because of eqs.
(\ref{gold1}) and (\ref{gibbs0}), only one parameter among the map
weights and contraction rates is left free, and can be put in one
to one relation with $D_0$.

Moreover, the two affine constants $\theta_i$ play no r\^ole in
determining the power--law behavior of the moments
$\nu_{\alpha}(t)$, since we can translate and stretch linearly the
support of $\mu$ with the only effect of multiplying the F-B.
functions by a complex number of modulus one, and of linearly
rescaling their argument \cite{except}. Notice finally that eq.
(\ref{gibbs0}) also implies that the I.F.S. is disconnected.

In conclusion, {\em the family of two--maps linear IFS measures
satisfying eq. (\ref{gold1}) can be partitioned into equivalence
classes labelled by the box--counting dimension $D_0(\mu)$}. The
distribution functions of three measures in the same equivalence
class are displayed in Fig. \ref{distr}. These measures enjoy
distinctive properties. First of all, they are {\em uniform Gibbs
measures}, according to the theory of Bowen \cite{bow}. Moreover,
since  eq. (\ref{ev452b}), with $\delta_i$ and $\pi_i$ in place of
$l_\sigma$ and $\pi_\sigma$, and $H(x)=1$ instead of the
asymptotic relation, defines the generalized dimensions of linear
I.F.S. measures {\em exactly}, one finds easily that
$D_q(\mu)=D_0(\mu)$ for {\em all} real values of $q$.

Now, figure \ref{zorro} shows the functions $\beta(\alpha)$
extracted numerically for three I.F.S. belonging to the same
equivalence class $D_0(\mu)=\frac{\log 2}{\log 5 - \log 2}$. The
coincidence of the curves (crosses) within numerical
precision---that we have also verified for other values of
$D_0$---lead us to {\em conjecture that the intermittency function
$\beta(\alpha)$ is an invariant of the equivalence classes defined
above}.

\section{From linear I.F.S. to potential theory: another conjecture}
Even prior to a formal proof of the conjecture just proposed,
accepting its validity leads to interesting speculations, and
raises intriguing questions. Clearly, the conjecture disproves any
relation of the kind $\beta(\alpha) = D_{q(\alpha)}$, with $q$ a
function of $\alpha$, like in the bounds (\ref{lowint}), or in the
Julia set relation (\ref{jul10}). This is not necessarily bad
news: it is just telling us once more that characteristics of the
measure $\mu$ other than the generalized dimensions determine the
dynamics: a few of these have been presented in this paper. How
are the exponents $d(r)$ and $d(\omega)$ of eqs. (\ref{term5}) and
(\ref{term5b}) related, in and across the equivalence classes
above? And the coefficients $\Gamma$ ? Moreover, we can also ask
how curves with different values of $D_0$ map among themselves.
But mostly, since eq. (\ref{gold1}) is magnificent in its
simplicity, is it there a simple argument to prove the conjecture?

\begin{figure}[ht]
\includegraphics[width=12cm,height=11cm]{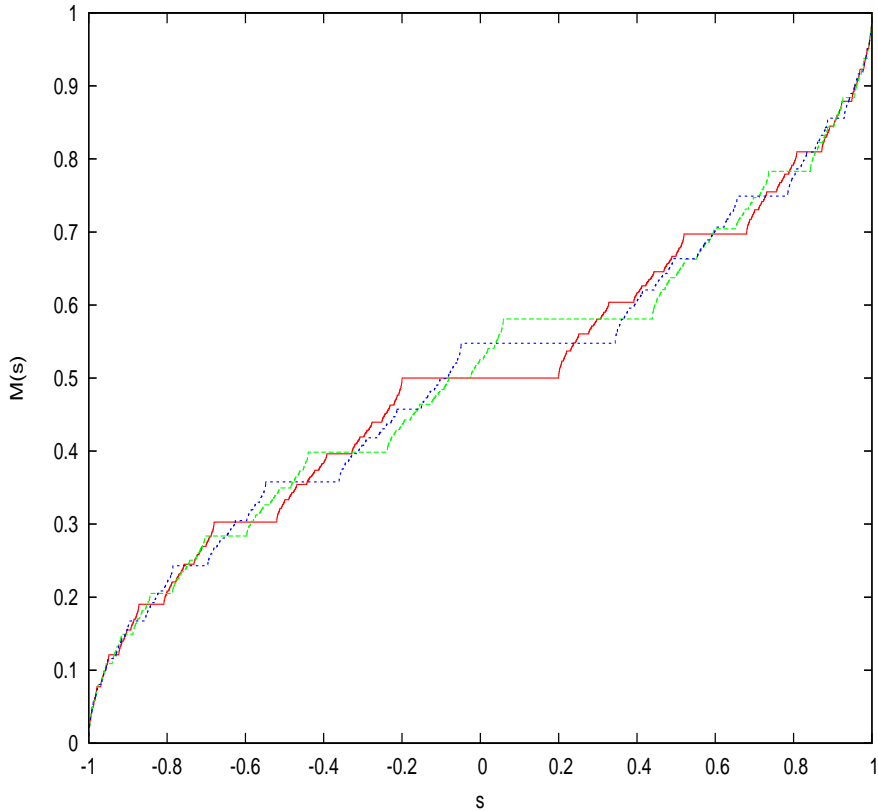}
\caption{Distribution functions $M(s):=\int_{-\infty}^s d\nu(s')$
where $\nu$ are the equilibrium measures associated with the three
I.F.S. measures of Fig. \ref{distr}.} \label{distr2}
\end{figure}

\begin{figure}[ht]
\includegraphics[width=12cm,height=9.5cm]{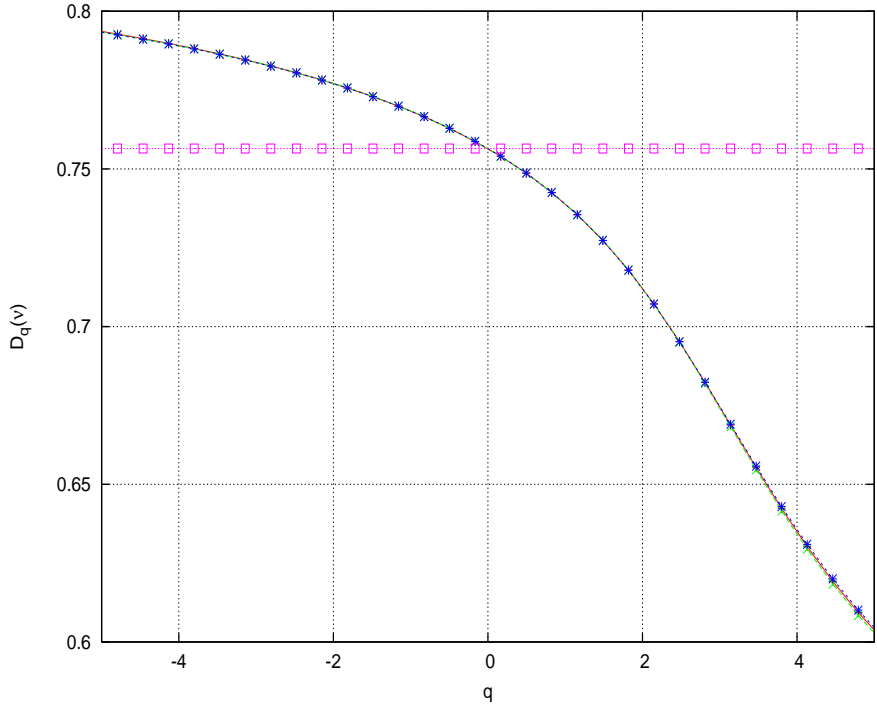}
\caption{Generalized dimensions of the three I.F.S. measures of
Fig. \ref{distr} (squares, horizontal line) and of the associated
equilibrium distributions (stars) of Fig. \ref{distr2}.}
\label{bernardo}
\end{figure}

{\em En suite}, notice that the extension of the conjecture to
I.F.S. with three or more maps, without further specifications, is
not valid. In fact, an instructive counter--example is obtained
setting $M=3$, and all contraction values and weights equal among
themselves: $\delta_i=\delta<\frac{1}{3}$, $\pi_i = \frac{1}{3}$,
for $i=1,2,3$. In this case, out of the three affine constants
$\theta_i$, two can be set arbitrarily (for instance, so that
$[-1,1]$ is the convex hull of the support of $\mu$), and one is
left free to vary. This can be done so that the resulting I.F.S.
is disconnected: its hierarchical structure is then composed of
the iteration of three {\em bands}, the position of the central of
which is variable.  The one-parameter set of I.F.S. measures so
obtained is composed of uniform Gibbs measures with
$D_q(\mu)=D_0(\mu)=-\log(\delta)/\log(3)$. And yet, the functions
$\beta(\alpha)$ are not invariant in this set: see figure
\ref{zorro}, where $\delta$ is chosen so to obtain the same $D_0$
as in the two--maps case.

We can try an explanation of this fact. These latter three--maps
I.F.S. measures have different quantum intermittency functions,
even if they coincide ``cylinderwise'', because their {\em
logarithmic potential equilibrium measures} are different, since
gaps between covering sets $I_\sigma$ have different geometric
ratios, and consequently, we expect different asymptotic behavior
of their orthogonal polynomials.

Having realized this, let us go back to the two--maps case. In
what respect are then the equilibrium measures within the
two--maps equivalence classes defined above ``equivalent''? Direct
inspection of their distribution functions, Fig. \ref{distr},
provides no clue. It is Fig. \ref{bernardo} that contains the
answer: {\em the generalized dimensions of the equilibrium
measures of two--maps I.F.S. in the same equivalence class are the
same}. One can therefore take the risk of putting forward a bolder
conjecture: {\em the intermittency function of uniform Gibbs
measures, whose equilibrium measures are characterized by the same
spectrum of generalized dimensions, are the same}
\cite{notagibbs}.

\section{Conclusions}
I have discussed in this paper a number of topics that have
originally been developed by mathematical physicists interested in
quantum mechanics, as it appears clearly from the list of
references, but that would certainly profit a lot from the
interest of specialists in orthogonal polynomials, special
functions, and potential theory. In fact, my formulation via
Fourier--Bessel functions, the original idea \cite{daniel,papgua}
to study these problems in relation with Jacobi matrices of
Iterated Function Systems, and my introduction of the
renormalization approach of orthogonal polynomials denote clearly
how much I owe to the community that has gathered for this
conference, and that I had the fortune to meet back in my postdoc
years here in Atlanta.

I have presented novel results on the asymptotic properties of F-B.
functions for Julia set invariant measures, relating different asymptotics
of the ``wave--packets'' $\nu_\alpha(N,\omega)$ to the properties
of the invariant measure, and of its orthogonal polynomials.
But mostly I have put forward open problems, numerical results, and
conjectures that indicate--I hope--where to search for complete answers.
How to turn this insight into a {\em constructive}
technique for determining the intermittency function is the job
that stays ahead. Having so arrived at the main topic of this
conference, potential theory and its applications, I can certainly
renew my best wishes to Ed, and retire in order.

\section*{Appendix: Numerical Techniques}
In addition to standard procedures, the research described in this
paper has required novel numerical algorithms for two main
problems: the construction of Jacobi matrices of linear I.F.S.,
described in \cite{cap}, \cite{mobius}, and the computation of the
F-B. functions \cite{oberw}.


\begin{thebibliography}{10}
\bibitem{negamome}
For $\alpha=0$, $\nu_0(t)=1$ is the normalization condition
(\ref{norma}). For negative values of $\alpha$ we can still define
moments by letting the summation in eq. (\ref{anu1}) start from
$n=1$. These moments are of interest when completing the analogy
with theory of the generalized dimension of singular measures.
\bibitem{notabd} In ref. \cite{ket}, various examples are exhibited for which
the bound $D_2(\mu)/d$ is significantly better than $D_1(\mu)$.
Yet, the (incorrect) statement is made that $D_2(\mu)/d$ is an
{\em upper} bound to $\beta(\alpha)$ for all negative values of
$\alpha$ (see \cite{negamome}). Where this to be true, it would
imply that $D_2(\mu)/d=\beta(0)$, where the latter quantity can be
defined by a limiting procedure, when $\beta(\alpha)$ is
continuous. To the contrary, in Sect. \ref{ssurf} we show a case
where $\beta(0)=D_1(\mu) > D_2(\mu)/d$.
\bibitem{notagibbs} Notice that Gibbs uniformity is required in this
conjecture. In fact, it is not enough to require that  measures be
characterised by equilibrium measures with the same generalized
dimensions. In fact, choose any of the two--maps I.F.S. of this
section, and change the weights $\pi_i$ while keeping the
contraction rates fixed. Since the equilibrium measure depends
only on the support of $\mu$, it does not change in the process.
To the contrary, as shown already in \cite{papgua}, in these
circumstances the intermittency functions are sensitive to the
weights.
\end{thebibliography}
\end{document}